\documentclass[]{aa} 

\usepackage{graphicx}
\usepackage{txfonts}
\def\vsini{\ensuremath{v \sin i}}

\begin{document}
\title{Photoelectric search for peculiar stars in open clusters. XV. Feinstein~1, NGC~2168, NGC~2323, NGC~2437, NGC~2547, 
NGC~4103, NGC~6025, NGC~6633, Stock~2, and Trumpler~2\thanks{Electronic version of the photometric data is only available at CDS via anonymous
ftp to cdsarc.u-strasbg.fr (130.79.128.5) or via
http://cdsarc.u-strasbg.fr/viz-bin/qcat?J/A+A/.../...}}

\author{E.~Paunzen\inst{1}
\and M.~Netopil\inst{2}
\and H.M.~Maitzen\inst{2}
\and K.~Pavlovski\inst{3}
\and A.~Schnell\inst{2}
\and M.~Zejda\inst{1}}
\institute{Department of Theoretical Physics and Astrophysics, Masaryk University,
Kotl\'a\v{r}sk\'a 2, 611\,37 Brno, Czech Republic \\
\email{epaunzen@physics.muni.cz}
\and
Universit{\"a}tssternwarte, T{\"u}rkenschanzstr. 17, A-1180 Wien, Austria
\and
Department of Physics, Faculty of Science, University of Zagreb, Bijeni\v{c}ka cesta 32, 10 000, Zagreb, Croatia}

   \date{} 
  \abstract
   {The chemically peculiar (CP) stars of the upper main sequence are mainly characterized by strong overabundances
	  of heavy elements. Two subgroups (CP2 and CP4) have strong local magnetic fields which make them interesting
		targets for astrophysical studies. This star group, in general, is often used for the analysis of stellar formation
		and evolution in the context of diffusion as well as meridional circulation.}
   {In continuation of a long term study of CP stars (initiated in the 1980ies), we present new results based on photoelectric 
	  measurements for ten open clusters that are, with one exception, younger than 235\,Myr. 
		Observations in star clusters are favourable because they represent samples of stars of constant age and homogeneous 
		chemical composition.}
   {The very efficient tool of $\Delta a$ photometry was applied. It samples the flux depression at 5200\AA\,
	  typically for CP stars. In addition, it is able to trace emission line Be/Ae and $\lambda$ Bootis stars. Virtually all CP2 and CP4 stars 
		can be detected via this tool, and it has been successfully applied even in the Large Magellanic Cloud. For all targets in the cluster
		areas, we performed a kinematic membership analysis.}
	 {We obtained new photoelectric $\Delta a$ photometry of 304 stars from which 207 objects have a membership probability higher than 50\%.
	  Our search for chemically peculiar objects results in fifteen detections. The stars have masses between 1.7\,M$_{\sun}$ and 7.7\,M$_{\sun}$
		and are between the zero- and terminal-age-main-sequence. We discuss the published spectral classifications in the light of 
		our $\Delta a$ photometry and identify several misclassified CP stars. We are also able to establish and support the 
		nature of known bona fide CP candidates.}
   {It is vital to use kinematic data for the membership determination and also to compare published spectral types
	  with other data, such as $\Delta a$ photometry. There are no doubts about the accuracy of photoelectric measurements, especially for stars brighter than 10th magnitude. The new and confirmed CP stars are interesting targets for spectroscopic follow-up observations
		to put constraints on the formation and evolution of CP stars.}

\titlerunning{Photoelectric search for peculiar stars in open clusters. XV.}
\authorrunning{Paunzen et al.}
\maketitle

\section{Introduction}

More than a century ago, \citet{Maur97} detected a subclass of
A-type stars with peculiar lines and line strengths, which thereafter
became known as Ap stars. Later on, the spectral range was widened and
the class become known as chemically peculiar (CP) stars of the upper
main sequence. These stars revealed other peculiar features, for example 
the existence of a strong global magnetic field (CP2 and CP4 objects) 
with a predominant dipole component located at random with respect to the
stellar rotation axis and the centre of the star as well as overabundances
with respect to the Sun for heavy elements such as
silicon, chromium, strontium, and europium. 
The peculiar surface abundances for CP stars have been
explained either by diffusion of chemical
elements depending on the balance between gravitational
pull and uplift by the radiation field through absorption in spectral
lines or by selective accretion from the interstellar
medium via the stellar magnetic field \citep{Szkl13}. Therefore, the correlation of stellar magnetic field strengths with astrophysical processes like 
diffusion and meridional circulation as well as their evolutionary status 
can be very well studied with this stellar group \citep{Glag13}.

\begin{table*}
\caption{Fundamental parameters of the target clusters taken from \citet{Paun06} and \citet{Zejd12}.}
\label{cluster_par}
\begin{center}
\begin{tabular}{lcccrrrcrrr}
\hline
\hline 
Cluster & & $\alpha$(2000) & $\delta$(2000) & \multicolumn{1}{c}{$l$} & \multicolumn{1}{c}{$b$} & $d_{\sun}$ & $E(B-V)$ &  \multicolumn{1}{c}{$age$} & \multicolumn{1}{c}{$\mu_{\alpha}\cos\delta$} & \multicolumn{1}{c}{$\mu_{\delta}$} \\
        & &                &                &                         &                         & [pc]       & [mag]    & [Myr]                      & [mas/yr]                                     & [mas/yr]       \\
\hline
Feinstein~1 & C1103$-$595 & 11 05 56 & $-$59 49 00 & 290.03 & +0.39   & 1180 & 0.41 &   5 & $-$6.1 & +2.9    \\
NGC~2168    & C0605+243   & 06 09 00 & +24 21 00   & 186.59 & +2.22   & 830  & 0.23 & 100 & +1.5   & $-$2.9  \\
NGC~2323    & C0700$-$082 & 07 02 42 & $-$08 23 00 & 221.67 & $-$1.33 & 895  & 0.23 & 100 & +0.4   & $-$2.0  \\
NGC~2437    & C0739$-$147 & 07 41 46 & $-$14 48 36 & 231.86 & +4.06   & 1495 & 0.16 & 235 & $-$5.0 & +0.4    \\
NGC~2547    & C0809$-$491 & 08 10 09 & $-$49 12 54 & 264.47 & $-$8.60 & 430  & 0.05 &  45 & $-$7.7 & +5.0    \\
NGC~4103    & C1204$-$609 & 12 06 40 & $-$61 15 00 & 297.57 & +1.16   & 1810 & 0.25 &  30 & $-$5.6 & $-$0.5  \\
NGC~6025    & C1559$-$603 & 16 03 17 & $-$60 25 54 & 324.55 & $-$5.88 & 725  & 0.28 &  75 & $-$3.3 & $-$3.1  \\
NGC~6633    & C1825+065   & 18 27 15 & +06 30 30	 & 36.01  & +8.33   & 335  & 0.18 & 505 & +0.2   & $-$1.2  \\
Stock~2     & C0211+590   & 02 15 00 & +59 16 00	 & 133.33 & $-$1.69 & 300  & 0.33 & 130 & +16.6  & $-$13.5 \\
Trumpler~2  & C0233+557   & 02 37 18 & +55 59 00   & 137.38 & $-$3.97 & 605  & 0.32 & 120 & +1.0   & $-$4.6  \\
\hline
\end{tabular}
\end{center}
\end{table*}

\begin{table*}
\caption{Observations log, the description of the used equipment can be found in the last column.}
\label{obs_log}
\begin{center}
\begin{tabular}{lcccc}
\hline
\hline 
Cluster & Observatory & Telescope & Time & Reference \\
\hline
Feinstein~1 & ESO  & Bochum 0.61\,m & 89/04               & \citet{Mait93}  \\
NGC~2168    & Hvar & 0.65\,m        & 89/01, 89/02        & \citet{Mait87b} \\
NGC~2323    & ESO  & 0.5\,m         & 85/02, 85/03        & \citet{Mait87a} \\
NGC~2437    & ESO  & 1.0\,m         & 84/02               & \citet{Mait93}  \\
NGC~2547    & ESO  & 1.0\,m         & 84/02               & \citet{Mait93}  \\
NGC~4103    & ESO  & 1.0\,m         & 84/02               & \citet{Mait93}  \\
NGC~6025    & ESO  & Bochum 0.61\,m & 89/04               & \citet{Mait93}  \\
NGC~6633    & ESO  & Bochum 0.61\,m & 89/04               & \citet{Mait93}  \\
Stock~2     & Hvar & 0.65\,m        & 84/12, 85/12        & \citet{Mait87b} \\
Trumpler~2  & Hvar & 0.65\,m        & 85/12, 86/10, 86/11 & \citet{Mait87b} \\
\hline
\end{tabular}
\end{center}
\end{table*}

Nearly four decades ago, \citet{Mait76} introduced the $\Delta a$ 
photometric system in order to investigate the flux depression at 5200\AA\,
typically for CP stars. An overview of the system and its applications
can be found in \citet{Paun05}. The $a$ index samples the depth of this 
flux depression by comparing the flux at the centre with the adjacent regions.
The final intrinsic peculiarity index $\Delta a$ was defined as the difference between the individual 
$a$-values and the $a$-values of non-peculiar stars of the same colour (spectral type).
It was shown \citep{Paun05} that virtually all CP2 and CP4 stars have positive
$\Delta a$-values up to 95\,mmag. Extreme cases of the non-magnetic CP1 and CP3 objects 
may exhibit marginally positive $\Delta a$ values, whereas emission line Be/Ae and 
$\lambda$ Bootis stars exhibit significant negative values. Since the detailed study of 
Pleione \citep{Pavl89}, it is known that Be stars can change their $\Delta a$ values 
from significantly positive at their shell to negative at their emission phase. 

Starting with the paper by \citet{Mait81}, fourteen parts of a large photoelectric
$\Delta a$ survey to detect CP stars in open clusters and stellar associations
were published. Observations in star clusters are preferable because they represent 
samples of objects of constant age and homogeneous chemical 
composition, suited to the study of processes linked to stellar structure and 
evolution, and to fixing lines or loci in several very important astrophysical diagrams such 
as the colour-magnitude diagram (CMD), or the Hertzsprung-Russell diagram (HRD). 

In this paper, we present new photoelectric $\Delta a$ data for the ten open clusters 
Feinstein~1, NGC~2168, NGC~2323, NGC~2437, NGC~2547, 
NGC~4103, NGC~6025, NGC~6633, Stock~2, and Trumpler~2. All aggregates, with the exception of NGC~6633, 
are younger than 250\,Myr
(Table \ref{cluster_par}). The detection of CP stars in young open clusters will
help us to understand the formation and evolution of these objects and their magnetic fields.

\begin{figure*}
\begin{center}
\includegraphics[width=165mm]{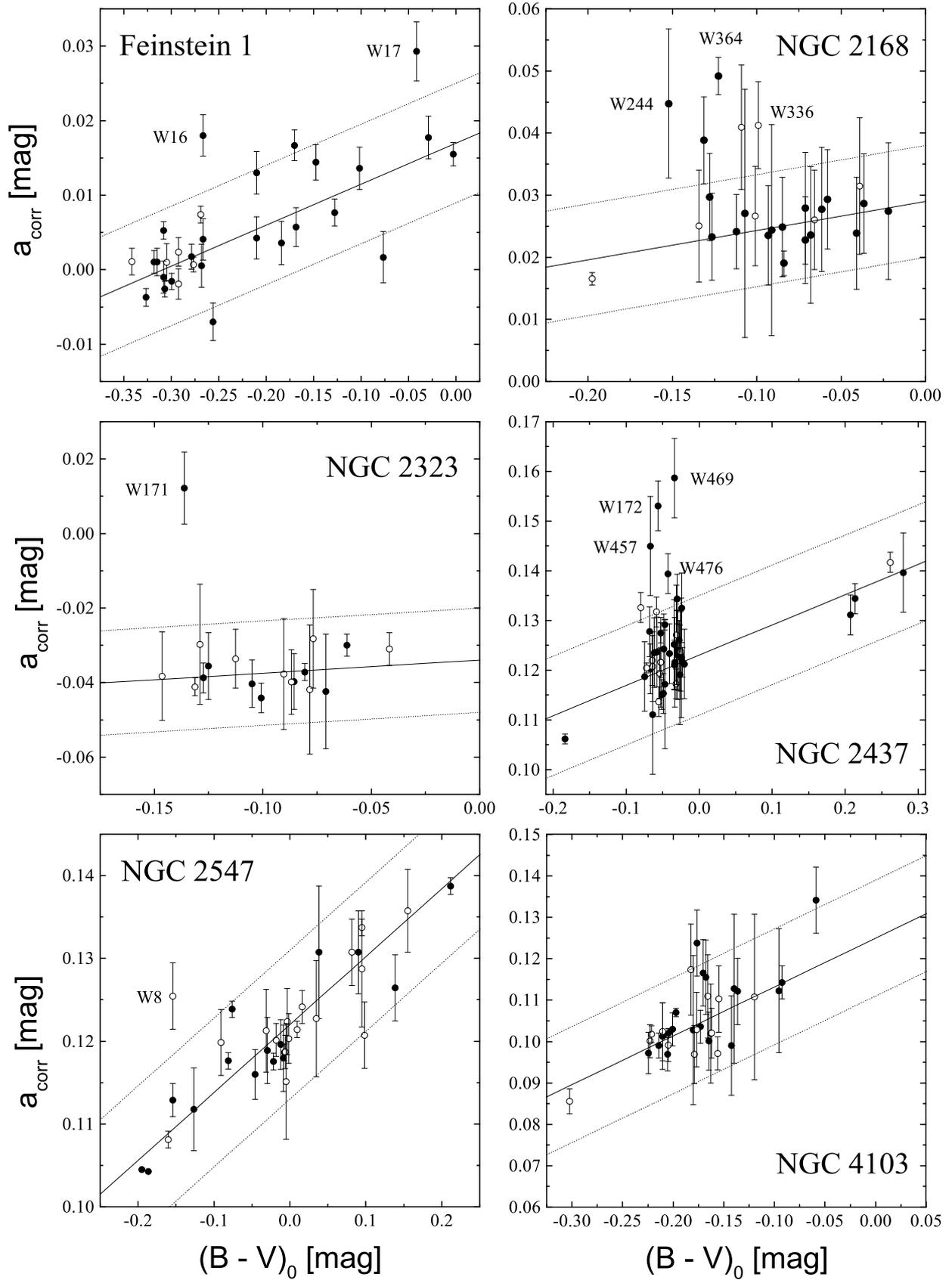}
\caption{$a_{\rm corr}$ versus $(B-V)_0$ diagrams for target clusters. Filled circles
denote stars with a kinematical membership probability of more than 50\%, open circles less than
50\%. Stars with a statistical
significant $\Delta a$ value, are denoted with their WEBDA numbers (W no.). The solid line is the
normality line and the dotted lines are the confidence intervals corresponding to 99.9\,\%.}
\label{aversusbv}
\end{center}
\end{figure*}
\addtocounter{figure}{-1}
\begin{figure*}
\begin{center}
\includegraphics[width=165mm]{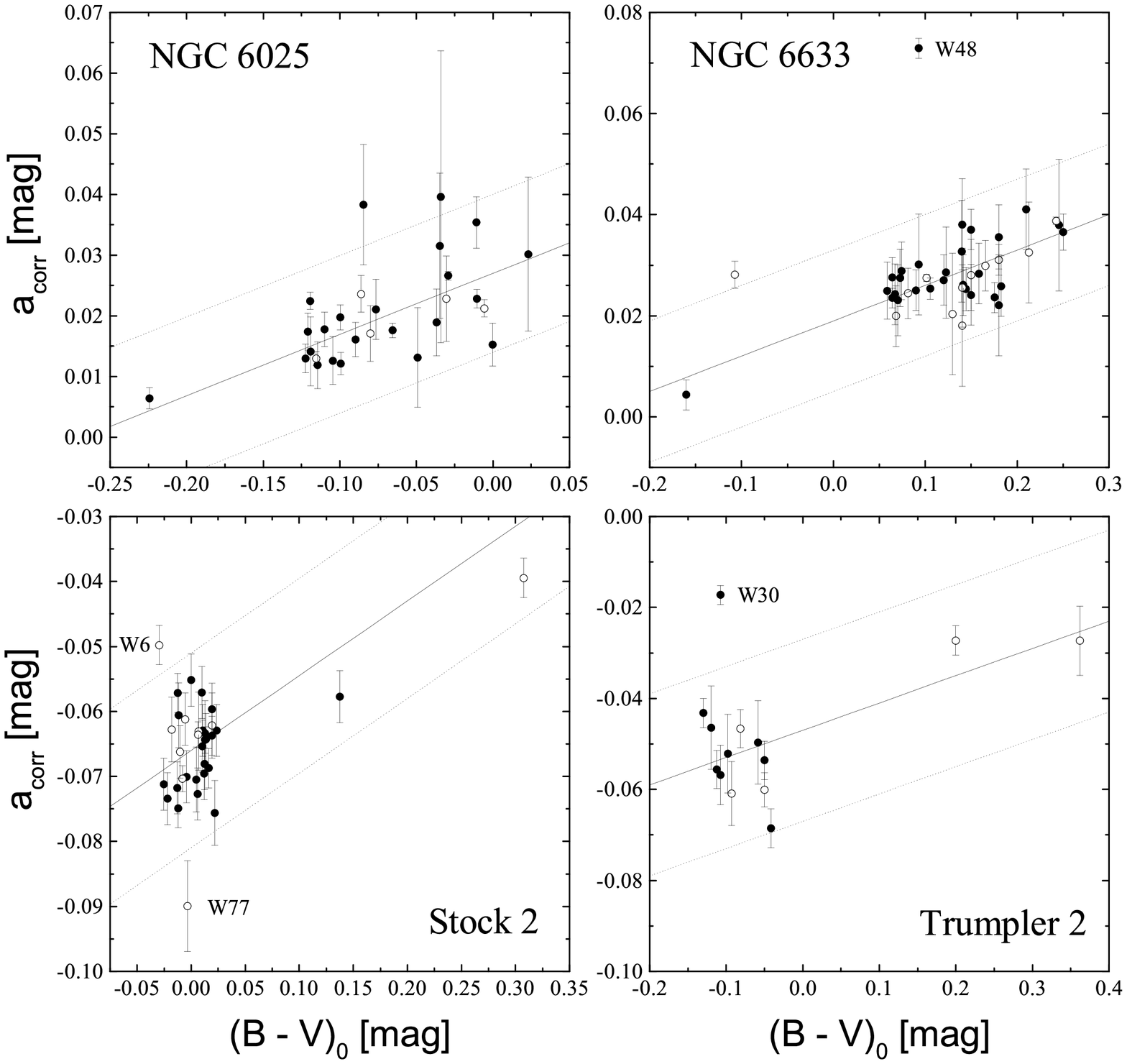}
\caption{continued.}
\label{aversusbv}
\end{center}
\end{figure*}

\section{Target selection, observations, and reduction} 

This work was initiated by the European working group on chemically peculiar stars of the upper main sequence \citep{Math89}.
The regular photoelectric observations of this programme were performed until 1991. Soon afterwards, the, for that time new,
technique of CCD photometry was employed \citep{Mait97}. Nevertheless, there are no doubts about the accuracy of the photoelectric measurements.
Furthermore, we also have time series of CP stars in our archive, which are very important for filling gaps when determining
rotational periods \citep{Miku11}. These data will be published in a separate paper.

We have chosen ten open clusters for which no $\Delta a$ observations
are yet available. The fundamental parameters of the target clusters taken from \citet{Paun06} and \citet{Zejd12} are listed in
Table \ref{cluster_par}. The observations log is listed in Table \ref{obs_log}. The details of the used equipment and basic
reduction processes can be found in \citet{Mait87a}, \citet{Mait87b}, and \citet{Mait93}.

To deredden the programme stars, we made use of photometric data in the Johnson, Geneva, and Str{\"o}mgren systems, compiled from 
the open cluster database WEBDA\footnote{http://webda.physics.muni.cz}.
For the first two systems, the well-known calibrations based on the X/Y parameters \citep{Cram99}
and Q-index \citep{Guti75}, respectively, applicable for O/B-type
stars were applied. Objects with available Str{\"o}mgren data were treated with the routines by \citet{Napi93}, allowing
the dereddening of cooler type stars. For member stars without sufficient photometry, the mean cluster reddening was adopted. 
There are only a few non-member stars (fifteen in total), for which we were not able to derive individual reddening values; these were therefore 
rejected in the subsequent analysis. If several estimates for a particular object are available, a mean value was calculated.
Since all clusters are rather near to the Sun, a strong reddening especially for cool-type stars is hardly expected.

Since the $a$ index is slightly dependent on 
temperature (increasing towards lower temperatures), the intrinsic peculiarity 
index $\Delta a$ had to be defined as the difference between the individual 
$a$-values and those of non-peculiar stars ($a_{\rm 0}$) of the 
same colour. The locus of the $a_{\rm 0}$-values has been called the normality line.
Because of the reddening, the normality line is shifted by $E(B-V)$
to the red and by a small amount $E(a)$ to higher $a$-values \citep{Mait93}.
The mean ratio of these shifts $f\,=\,E(a)/E(B-V)$ was determined as 
$f\,\approx\,0.035$. The final values were calculated as $a_{\rm corr} = a({\rm obs}) - fE(B-V)$.
For the subsequent analysis, we used the dereddened diagnostic $a_{\rm corr}$ versus $(B-V)_0$ diagrams (Fig. \ref{aversusbv}).
The mean $(B-V)$ values have been calculated from the data included in WEBDA.

The normality line for each cluster was determined using the photometric data of member and non-member stars, which is justified because of the use of individually dereddened colours.
Objects deviating more than 3\,$\sigma$ were rejected in an iterative process. The final 
coefficients of $a_{\rm corr} = b + c (B-V)_0$ together with their errors are listed in Table \ref{results}.
We consider an object as positively (or negatively) detected if its $\Delta a$ value, taking into account the
observational error, lies above (or below) the 3\,$\sigma$ limit of the corresponding normality line.

The $\Delta a$, $a_{\rm corr}$, and $(B-V)_0$ data of all stars (Fig. \ref{aversusbv}) are available from the
first author and/or via CDS. In the following, we will use the numbering system from WEBDA (W no.).

\begin{table*}
\caption{Final results, all photometric values are given in mmag. For Feinstein~1 W17 and Stock~2 W77, we were not able to derive reliable 
astrophysical parameters (see text). The errors in the final digits of
the corresponding quantity are given in parentheses.}
\label{results}
\begin{center}
\begin{tabular}{lccccc}
\hline
\hline 
                                   & Feinstein~1       & NGC~2168         & NGC~2323           & NGC~2437        & NGC~2547             \\
\hline
$a_{\rm corr} = b + c (B-V)_0, N$  & 17(2)/55(6)/24    & 29(2)/47(17)/21  & $-$34(4)/35(17)/17 & 123(1)/61(6)/37 & 122(1)/82(6)/29      \\
3\,$\sigma$                        & $\pm$8            & $\pm$9           & $\pm$14            & $\pm$12         & $\pm$9               \\
$N(>50\%)/N(<50\%)$                & 22/6              & 19/8             & 9/9                & 36/12           & 16/17                \\
\#CP/$\Delta a$/Prob               & W16/+16/99        & W244/+16/93      & W171/+51/82        & W172/+34/80     & W8/+16/8             \\
                                   & W17/+15/59        & W336/+17/37      &                    & W457/+26/76     &                      \\
																	 &                   & W364/+23/69      &                    & W469/+38/63     &                      \\
																	 &                   &                  &                    & W476/+19/50     &                      \\
$\log T_{\rm eff}/\log L/L_{\sun}$ & 4.350/3.36        & 4.126/2.32       & 4.105/2.07         & 3.993/1.67      & 4.124/2.245          \\
                                   & $-$/$-$           & 4.060/2.25       &                    & 4.018/1.82      &                      \\
                                   &                   & 4.086/2.77       &                    & 3.978/1.98      &                      \\
                                   &                   &                  &                    & 3.985/1.72      &                      \\
\hline
                                   & NGC~4103          & NGC~6025         & NGC~6633           & Stock~2           & Trumpler~2         \\
\hline
$a_{\rm corr} = b + c (B-V)_0, N$  & 129(5)/151(25)/34 & 27(2)/101(16)/24 & 19(2)/70(11)/38    & 125(4)/118(21)/31 & $-$47(2)/60(14)/12 \\
3\,$\sigma$                        & $\pm$14           & $\pm$13          & $\pm$14            & $\pm$15           & $\pm$20            \\
$N(>50\%)/N(<50\%)$                & 21/13             & 23/5             & 28/12              & 24/10             & 9/5                \\
\#CP/$\Delta a$/Prob               &                   &                  & W48/+47/65         & W6/+20/0          & W30/+42/66         \\
                                   &                   &                  &                    & W77/$-$24/4       &                    \\
$\log T_{\rm eff}/\log L/L_{\sun}$ &                   &                  & 3.919/0.92         & 3.988/1.49        & 4.069/2.12         \\
                                   &                   &                  &                    & $-$/$-$           &                    \\
\hline
\end{tabular}
\end{center}
\end{table*}

For the membership probabilities of the individual stars, we employed the method given in \citet{Bala98}. In addition,
we compared the results with those of the algorithm published by \citet{Java06}, yielding excellent agreement. The first
method takes both the errors of the mean cluster and the stellar proper motions into account.
The mean proper motions of the target clusters were taken from \citet{Zejd12} and are listed in Table \ref{cluster_par}. The
proper motions of the individual stars were taken from the following sources, sorted by the priority:
\begin{itemize}
\item TYCHO-2 \citep{Hog00};
\item UCAC4 \citep{Zach13};
\item PPMXL \citep{Roes10}.
\end{itemize}
For the complete sample, kinematic data are available. Because the mean proper motions of NGC~2547 and
Stock~2 deviate strongly from the background, a definite membership determination should be straightforward.
A comparison with the results published by \citet{Baum00} yields an excellent agreement.

To place the detected CP candidates in the HRD (Fig. \ref{hrd}), we 
applied the effective temperature calibrations and bolometric corrections for CP stars by 
\citet{net08} on the available photometric data in the Johnson, Geneva, and Str{\"o}mgren systems. 
The luminosities were derived using the cluster parameters listed in Table \ref{cluster_par} regardless 
of the kinematic non-membership in order to obtain an additional membership criterion. In general, we adopt an uncertainty for 
the derived temperatures of 500\,K and 700\,K for CP2 and CP4 stars, respectively \citep[see][]{net08}. The errors in 
luminosity were derived using a standard error in distance of 10\,\%, 0.1\,mag for bolometric correction, and 0.02\,mag 
for brightness and interstellar reddening.

By comparing the derived absolute magnitudes and temperatures with available 
spectral types, and the position of the stars in the HRD with the adopted cluster ages, 
we conclude that all kinematic non-members, except NGC~2547 W8, can be also considered  
cluster non-members from the photometric point-of-view.

One exceptional case is Feinstein~1 W17 which is a kinematical member, but from
photometry we conclude the contrary. A detailed analysis for it is given in
Sect. \ref{feinstein1}.

\section{Discussion}

From our final $a_{\rm corr}$ versus $(B-V)_0$ diagrams (Fig. \ref{aversusbv}), we find fifteen stars
that deviate significantly from the corresponding normality lines. About one third of them seem not
to be members of the associated star cluster. However, since we have dereddened each star individually,
these stars are still very good candidates for being true CP stars. In the following, we discuss the 
results for the open clusters in more detail.

\subsection{Feinstein~1} \label{feinstein1}

This aggregate is an accumulation of brighter stars around the $\beta$ Cephei variable and helium star 
HD~96446 (W20). If this is indeed a true star cluster, it is very young and all A- and F-type stars should 
still be in there Pre-Main-Sequence phase \citep{Fein64}. Since its initial discovery, it has hardly been studied.

We find two stars, HD~305941 (W16) and HD~306034 (W17), significantly above the normality line.
\citet{Garc93} classified W16 as B2 IV/V (\vsini\ of 80\,km\,s$^{-1}$) 
and W17 as Am (kA2\,IV, mF6\,III/IV). The projected rotational velocity of W16 is exceptionally
low for an early B-type star which already points to a CP nature. They also noted that
``\ion{He}{II} at 4009\AA\ presents intensity variations''. From the classification of W17 it is already 
clear that this is a very extreme CP1 star. 

From the kinematic data, we derive membership probabilities of 99\% and 59\%, respectively.
Therefore, we calibrated these stars using the reddening and distance of Feinstein~1 (Table \ref{cluster_par}). 
This resulted in $T_{\rm eff}$ and $M_{\rm Bol}$ of [22400\,K, $-$3.7\,mag] and [10300\,K, $-$1.6\,mag] for 
W16 and W17. While the values for W16 are perfectly in line with the spectral type, they are
not compatible for W17 even if we assume a pre-main-sequence 
status. For W17, we chose a different approach on the basis of Johnson $UBV$ photometry. Taking the
standard relation for $(B-V)$ versus $(U-B)$ from \citet{Schm82} and the observed values, we estimated
the reddening assuming that the object follows the standard relation. As a result, we get $E(B-V)$\,=\,0.15\,mag. 
With this reddening value, we calibrated the $T_{\rm eff}$ as 7800\,K.
This value is well in line with the spectral type (Am (kA2\,IV, mF6\,III/IV)). As a last step, we kept the $T_{\rm eff}$ value fixed and 
calculated the distances which place the star either at the zero- or terminal-age-main-sequence within
the evolutionary grids by \citet{Scha92}. The derived values are between 300\,pc and 850\,pc which establishes 
it as being a foreground star. Recalculating the $\Delta a$ value using the reddening given above and 
the relation from Table \ref{results} yields +10\,mmag. However, this peculiar object is worthwhile for follow-up spectroscopic 
observations. We conclude that W17 is not associated with Feinstein~1, although its kinematic characteristic is, to a certain degree,
compatible with the mean cluster proper motion. 
 
From the location in the HRD, we infer that W16 is a CP4 (He-strong) star with about 7.7\,M$_{\sun}$. 
This star is the most massive and most luminous star among our CP candidates (Fig. \ref{hrd}). \citet{Cida07}
presented a similar diagram for a sample of CP4 stars (see Fig. 7 therein). Comparing both diagrams, we find that
this star is among the youngest CP4 objects known so far. We encourage follow-up observations in order to 
detect a possible stellar magnetic field.

\subsection{NGC~2168}

For this open cluster, two CP star candidates are listed in the literature \citep{Nied88}: W364 (HD~252405)
and W547 (HD~252459). The source of the corresponding information is not known to us. Both stars are
classified in \citet{Hoag65} as B6\,III: and A3\,V, respectively. The star W364 has a measured $\Delta a$ value
of +23\,mmag therefore, we confirm its peculiar nature. For W547, we get +17\,mmag, but the error of the mean
is too large to make it statistically significant. One could argue that this is caused by an apparent photometric
variability due to spots and rotation. However, according to the kinematical data, W547 is not a member (1\% probability)
of NGC~2168. On the other hand, W364 is a member of the cluster. It is close to the terminal-age-main-sequence with a mass of
about 4.1\,M$_{\sun}$. This would correspond to a probable B7 Si-type star.

In addition, two other stars from our sample are significantly above the normality line, namely W244 
(TYC\,1877$-$356$-$1, +16) and W336 (HD~252427, +17). For neither of these objects are spectral classifications available \citep{Skif13}. The kinematical data supports the membership of W244 (93\%), but not for W336 (37\%).
They both have masses between 3\,M$_{\sun}$ and 4\,M$_{\sun}$ and would be therefore classified as late B-type Si stars. 
We note that HD~252427 is incorrectly identified as W335 in SIMBAD/CDS. For W336, the star HD~252458 is listed.

The only known emission-type object in the cluster area \citep{Koho99}, HD~41995 (W781), was not
measured.

\subsection{NGC~2323}

The known CP candidate stars within NGC~2323 are HD~52965 (W3) and BD$-$08\,1708 (W51). \citet{Bych09} listed upper limits of
the magnetic fields for these objects of 89\,G and 49\,G, respectively. They list a spectral classification of 
B8 Si, taken from \citet{Rens09}, for HD~52965 which is probably wrong. The only found reference in
this respect is from \citet{Youn73} which lists ``B9 p: \ion{Si}{II} (4128/4130) slightly enhanced''.
We measured a $\Delta a$ value of +3\,mmag. This value together with the non-detection of a magnetic field 
suggests that this object is not a CP star. The same authors published the spectral type of 
``B9p:: \ion{Hg}{II} 3984\AA\,weak'' for BD$-$08\,1708. A classification which is not typical for a CP3 object.
However, we have not measured this star.

\begin{figure}
\begin{center}
\includegraphics[width=88mm]{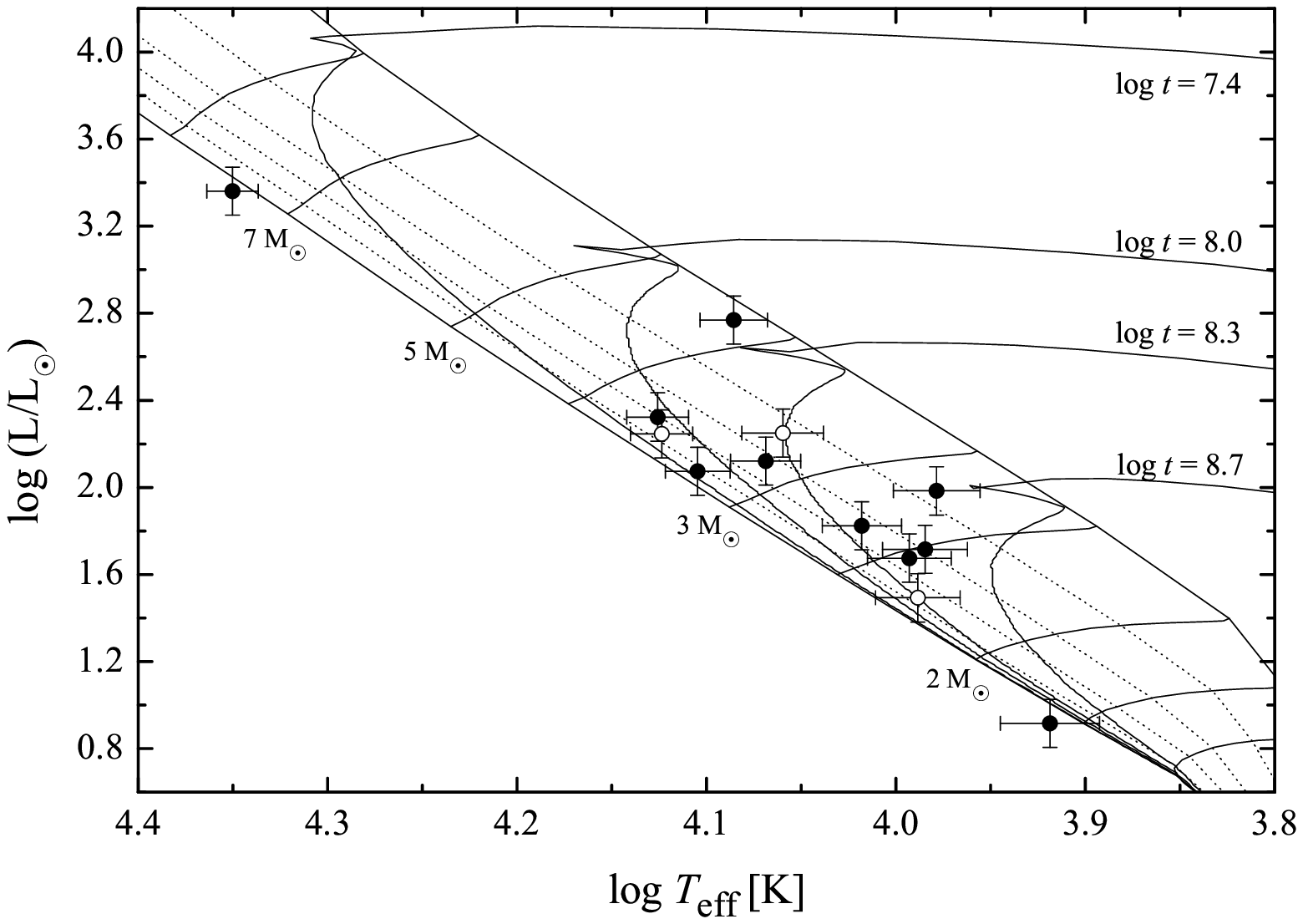}
\caption{Hertzsprung-Russell diagram for the bona fide CP stars listed in Table \ref{results}.
The dotted lines represent the lines of equal fractional ages calculated from the evolutionary grids by \citet{Scha92}. The stellar
mass, relative to each track, in units of solar masses, is indicated in the plot.
Both isochrones and evolutionary tracks are for solar metallicity. Kinematic non-members are represented by open symbols.}
\label{hrd}
\end{center}
\end{figure}

For the star CD$-$08\,1704 (W171) we find a significant positive $\Delta a$ value of +51\,mmag. Its membership
probability is 82\%. The are some recent investigations of NGC~2323 that include photometric but no 
spectroscopic data \citep{Clar98,Shar06,Frol12}. \citet{Clar98} list $E(B-V)$\,=\,0.23\,mag for W171 which corresponds
to the mean value of the cluster (Table \ref{cluster_par}). Its location within the HRD (Fig. \ref{hrd}) is consistent
with the isochrone of NGC~2323. The star W171 is therefore a 100\,Myr old, 3.4\,M$_{\sun}$ (late B-type), probable Si, star.

\subsection{NGC~2437}

The open cluster NGC~2437 has been studied photometrically several times because it might host a planetary nebula
\citep{Maja07}. However, there is no detailed spectroscopic analysis available in the literature. 
Neither CP nor Be stars within the cluster area have been detected so far.

We detected four stars that lie above the normality line, namely W172 (TYC\,5422$-$967$-$1, $\Delta a$\,=\,+34\,mmag),
W457 (TYC\,5422$-$2127$-$1, +26), W469 (BD$-$14\,2133, +38), and W476 (TYC\,5422$-$305$-$1, +19). For none of these objects are spectral classifications available in the literature. All stars are, within the errors, well represented by the isochrone for
NGC~2437 with a distance of 1500\,pc and an age of 235\,Myr. The derived masses range from 2.4\,M$_{\sun}$ to
2.9\,M$_{\sun}$ which corresponds to early A-types. The overall metallicity of NGC~2437 is solar
\citep{Paun10}. 

\subsection{NGC~2547}

The star HD~68074 (KW Vel, W8) is a known CP2 star with a rotational period of 1.1822\,d \citep{Nort87}.
Our measured $\Delta a$ value of +16\,mmag is slightly higher than that published by \citet{Mait83},
+9\,mmag. It is the only star measured by them in the cluster area. The values of the peculiar indices 
in the Geneva system \citep{Hauc82}, $\Delta(V1-G)$\,=\,+3\,mmag  
and $Z$\,=\,$-$12\,mmag, are intrinsically consistent. Because of the large amplitude of the rotational induced variability,
variations of the $\Delta a$ are also expected. From the current available kinematical data, we
deduce that W8 is not a member of NGC~2547 (probability of 8\%), a result that is in line with the
study by \citet{Baum00}, who listed a probability of 0\%. According to the different photometric diagrams,
it appears to be a member \citep{Clar82} because neither the cluster nor the star exhibits a reddening
larger than 0.05\,mag. Therefore, it is essential to use proper motions for a membership determination.

\citet{Nied88} listed HD~68275 (W35) as a CP1 star on the basis of the classification as A3(m)
by \citet{Houk78}. However, \citet{Hart76} listed a spectral type of A2\,V for this object. 
We find $\Delta a$\,=\,+2\,mmag. Even, if the CP1 nature can be confirmed, it is not a member of NGC~2547
on the basis of the proper motions (probability of 1\%). Therefore, NGC~2547 seems not to host any classical CP stars.

\subsection{NGC~4103}

The object CD$-$60\,3984 (W12) is a well-known emission-type star that could be still in the pre-main-sequence phase \citep{Heni76}.
We measured a $\Delta a$ value of $-$4\,mmag, which is not significant according to our criteria. From the
kinematical data, a membership probability of 49\% was calculated.

In total, we measured 34 stars in the cluster area, but found no significant deviating $\Delta a$ values 
for this young (30\,Myr) and rather metal-poor \citep[$-$0.47(10)\,dex; ][]{Paun10} aggregate.

\subsection{NGC~6025}

\citet{Tetz11} list an age of 40$\pm$11\,Myr for the emission-type star \citep{Heni76} HD~143448 (W1). Although their derived age is only marginally compatible with that of NGC~6025 (75\,Myr), but our derived 
membership probability is 94\%. It was therefore previously identified as a blue straggler by \citet{Merm82}.
The $\Delta a$ value of +2\,mmag for W1 is not significant. We did not measure CD$-$60\,6028, which was classified as A2\,Vp \citep{Paun01}.

Two objects, CD$-$60\,6011 (W13) and CD$-$60\,6002 (W38), have positive $\Delta a$ values but with large
errors. \citet{Gros11} listed the following spectral types and projected rotational velocity for these
two stars: B8\,V, 185\,km\,s$^{-1}$ and A0\,V. The high \vsini\ value of W13 is clearly not in line
with the values found for CP stars. If we assume that W38 is a bona fide CP star candidate, the large error of $a_{\rm corr}$ 
could be caused by rotational induced variability.

\subsection{NGC~6633}

\citet{Leva77} published spectral types of stars in the cluster area. Among their sample, they
identified five CP1 stars (W67, W75, W87, W88, and W161) and one CP2 star (W39). From these objects,
we measured four CP1 stars: W67 ($\Delta a$\,=\,$-$5\,mmag), W75 ($-$3), W87 (+3), and W88 ($-$8).
The non-detection of CP1 stars is consistent with previous results for this group \citep{Mait98}.
However, we are not able to draw any further conclusions from our measurements about the true nature
of these objects. 

\citet{Leva77} also identified HD~170054 (W77) as a blue straggler with a spectral type of B6\,IV. 
Later on, \citet{Abt85} changed the classification to B6\,IVp (Si). \citet{Bail13} used
this object as a ``standard star'' for determining \ion{Si}{II} and \ion{Si}{iii} elemental
abundances in B-type stars. They found no significant deviation from the solar abundance for it.
We measured $\Delta a$\,=\,$-$3\,mmag, 
which is consistent with the values of peculiar indices of the Geneva system, $\Delta(V1-G)$\,=\,$-$16 
and $Z$\,=\,+5\,mmag. We conclude that this object is not chemically peculiar and that it has been misclassified. 

There is one star, HD~169959 (W58), that is just on the limit of the detection level. It is a non-member
and has been classified as A0\,III by several authors \citep{Abt85}.

We find a significantly positive $\Delta a$ of +47\,mmag for star W48 (BD+06\,3755). In terms of
kinematic data, it is a member of NGC~6633. Unfortunately, no further information other than $uvby\beta$ photometry
is published in the literature. Its location in the HRD (Fig. \ref{hrd}) suggests that it is, within the errors,
very close to the zero-age-main-sequence with a mass of about 1.7\,M$_{\sun}$. This would correspond to a rather
cool late A-type CP2 star, probably with enhancements of chromium, strontium, and europium. 
Such objects are rare \citep{Poeh05} and are very important for the study of the early formation and evolution
of the local stellar magnetic field.

\subsection{Stock~2}

There are two CP stars listed in the catalogue by \citet{Rens09}, namely HD~13402 (W34) and
HD~13412 (CP1, W31). The first star is a high-mass supergiant (B0.5Iab) with a very small proper motion
and is probably misclassified \citep{Skif13}. The proper motion of HD~13412 ($-$14.73/+3.96) also disagrees with the mean 
cluster motion. As a consequence, \citep{Baum00} concluded that both stars are non-members. However, we did not measure these objects.

The star HD~13112 (W6) is catalogued as a blue straggler by \citet{Ahum95} and was classified as B8\,III (sharp hydrogen lines) 
by \citet{Mccu74}. The latter could be a hint of a CP nature because of the generally low
rotational velocities of these objects which were often misidentified as sharp lined giants \citep{Pres74}.
The location in the HRD (Fig. \ref{hrd}) is clearly in favour of it being on the main sequence. The
derived astrophysical parameters (Table \ref{results}) are typical for an A2-type star, whereas $(B-V)_0$\,=\,$-$0.03\,mag
is consistent with a B8-type object. Again, this irregular behaviour is typical of CP stars.
The final decision about its true nature can only be drawn on the basis of further observations, for
example, classification resolution spectroscopy.

For HD~13631 (W77) we find $\Delta a$\,=\,$-$24\,mmag, which is the only significant negative value in
the complete sample presented in this paper. In the literature, two different spectral classifications,
namely B9\,V \citep{Voro85} and A1\,II: \citep[hydrogen lines sharp;][]{Mccu74} have been published. 
Based on the colours of this star and the absolute magnitude of a class II bright giant, this object 
would be located at a distance of about 2\,kpc. This would imply a much higher interstellar reddening and a 
proper motion much closer to zero. We therefore conclude that the classification as B9\,V is more likely. 
However, both spectral types and proper motion contradict membership. Since \citet{Voro85} has not noticed 
any emission in the spectra, we suggest that W77 is a good candidate for being a hot $\lambda$ Bootis-type star.

\subsection{Trumpler~2}

The only known Be star in this cluster \citep{Schi76}, W8 (HD~16080), has a $\Delta a$ value of +7\,mmag.
It seems that it was measured in its shell phase. The star W30 (BD+55\,664) was
detected as a CP (Ap) star of Si-type by \citet{Zelw71} on the basis of prism 
spectroscopy. From kinematical data, it was confirmed as member of Trumpler~2
in the aforementioned paper. We derived $\Delta a$\,=\,+42\,mmag, which supports
the classification as a classical CP2 star. The recent and more accurate proper motions
establish this star as being a true member of Trumpler~2.

\section{Conclusions}

We presented new photoelectric $\Delta a$ photometry of 304 stars in ten open cluster
fields. From a detailed kinematical analysis, we concluded that 207 stars of the sample have
a membership probability higher than 50\%.

The ten clusters (Feinstein~1, NGC~2168, NGC~2323, NGC~2437, NGC~2547, 
NGC~4103, NGC~6025, NGC~6633, Stock~2, and Trumpler~2) are young to intermediate
age aggregates (5\,Myr\,$<$\,age\,$<$\,500\,Myr) with distances from 300\,pc to 1800\,pc from the Sun. 

Our search for chemically peculiar objects results in fifteen detections from which ten
objects seem to be true member of the corresponding star cluster. Of the remaining five stars, four are
probable field CP objects and one is $\lambda$ Bootis-type candidate. The objects have masses between 1.7\,M$_{\sun}$ 
and 7.7\,M$_{\sun}$ and are dwarf stars (luminosity class V).

We discussed the already published spectral classifications in the light of our $\Delta a$ photometry
and identified several misclassified CP stars. On the other hand, we were also able to establish and support
the nature of known bona fide CP candidates.

The newly detected CP stars, close to the zero-age-main-sequence, will help to understand the
formation and evolution of this phenomenon. To this end, follow-up observations, for example, to
detect and trace the local stellar magnetic fields, as well as an abundance analysis are needed.

\begin{acknowledgements}
This project is financed by the SoMoPro II programme (3SGA5916). The research leading
to these results has acquired a financial grant from the People Programme
(Marie Curie action) of the Seventh Framework Programme of EU according to the REA Grant
Agreement No. 291782. The research is further co-financed by the South-Moravian Region. 
It was also supported by the grants GA \v{C}R P209/12/0217, 14-26115P, 7AMB12AT003, 7AMB14AT015, and
the financial contributions of the Austrian Agency for International 
Cooperation in Education and Research (BG-03/2013, CZ-10/2012, and CZ-09/2014).
This research has made use of the WEBDA database, operated at the Department of 
Theoretical Physics and Astrophysics of the Masaryk University.
This work reflects only the author's views and the European 
Union is not liable for any use that may be made of the information contained therein.
\end{acknowledgements}


\begin{thebibliography}{}
\bibitem[\protect\citeauthoryear{Abt}{1985}]{Abt85} Abt, H. A. 1985, \apjl, 294, 103
\bibitem[\protect\citeauthoryear{Ahumada \& Lapasset}{1995}]{Ahum95} Ahumada, J., \& Lapasset, E. 1995, \aaps 109, 375
\bibitem[\protect\citeauthoryear{Bailey \& Landstreet}{2013}]{Bail13}	Bailey, J. D., \& Landstreet, J. D. 2013, \aap, 551, A30 
\bibitem[\protect\citeauthoryear{Balaguer-N\'unez et al.}{1998}]{Bala98} Balaguer-N\'unez, L., Tian, K. P., \& Zhao, J. L. 1998, \aaps, 133, 387
\bibitem[\protect\citeauthoryear{Baumgardt et al.}{2000}]{Baum00} Baumgardt, H., Dettbarn, C., \& Wielen, R. 2000, \aaps, 146, 251
\bibitem[\protect\citeauthoryear{Bychkov et al.}{2009}]{Bych09} Bychkov, V. D., Bychkova, L. V., \& Madej, J. 2009, \mnras, 394, 1338
\bibitem[\protect\citeauthoryear{Cidale et al.}{2007}]{Cida07} Cidale, L. S., Arias, M. L., Torres, A. F., et al. 2007, \aap, 468, 263
\bibitem[\protect\citeauthoryear{Clari{\'a}}{1982}]{Clar82} Clari{\'a}, J. J. 1982, \aaps, 47, 323
\bibitem[\protect\citeauthoryear{Clari{\'a} et al.}{1998}]{Clar98} Clari{\'a}, J. J., Piatti, A. E., \& Lapasset, E. 1998, \aaps, 128, 131
\bibitem[\protect\citeauthoryear{Cramer}{1999}]{Cram99} Cramer, N. 1999, \na, 43, 343
\bibitem[\protect\citeauthoryear{Feinstein}{1964}]{Fein64} Feinstein, A. 1964, Observatory, 84, 11
\bibitem[\protect\citeauthoryear{Frolov et al.}{2012}]{Frol12} Frolov, V. N., Ananjevskaja, Yu. K., \& Polyakov, E. V. 2012, Astronomy Letters, 38, 74
\bibitem[\protect\citeauthoryear{Garcia}{1993}]{Garc93} Garcia, B. 1993, \apjs, 87, 197
\bibitem[\protect\citeauthoryear{Glagolevskij}{2013}]{Glag13} Glagolevskij, Yu. V. 2013, Astrophysics, 56, 173
\bibitem[\protect\citeauthoryear{Grosso \& Levato}{2011}]{Gros11} Grosso, M., \& Levato, H. 2011, \rmxaa, 47, 255
\bibitem[\protect\citeauthoryear{Gutierrez-Moreno}{1975}]{Guti75} Gutierrez-Moreno, A. 1975, \pasp, 87, 805
\bibitem[\protect\citeauthoryear{Hartoog}{1976}]{Hart76} Hartoog, M. R. 1976, \apj, 205, 807
\bibitem[\protect\citeauthoryear{Hauck \& North}{1982}]{Hauc82} Hauck, B., \& North, P. 1982, \aap, 114, 23
\bibitem[\protect\citeauthoryear{Henize}{1976}]{Heni76} Henize, K. G. 1976, \apjs, 30, 491
\bibitem[\protect\citeauthoryear{Hoag \& Applequist}{1965}]{Hoag65} Hoag, A. A., \& Applequist, N. L. 1965, \apjs, 12, 215
\bibitem[\protect\citeauthoryear{H{\o}g et al.}{2000}]{Hog00} H{\o}g, E., Fabricius, C., Makarov, V.~V., et al. 2000, \aap, 355, L27
\bibitem[\protect\citeauthoryear{Houk}{1978}]{Houk78} Houk, N. 1978, Michigan catalogue of two-dimensional spectral types for the HD stars, University of Michigan
\bibitem[\protect\citeauthoryear{Javakhishvili et al.}{2006}]{Java06} Javakhishvili, G., Kukhianidze, V., Todua, M., \& Inasaridze, R. 2006, \aap, 447, 915
\bibitem[\protect\citeauthoryear{Kohoutek \& Wehmeyer}{1999}]{Koho99} Kohoutek, L., \& Wehmeyer, R. 1999, \aaps, 134, 255
\bibitem[\protect\citeauthoryear{Levato \& Abt}{1977}]{Leva77} Levato, H., \& Abt, H. A. 1977, \pasp, 89, 274
\bibitem[\protect\citeauthoryear{Maitzen}{1976}]{Mait76} Maitzen, H. M. 1976, \aap, 51, 223
\bibitem[\protect\citeauthoryear{Maitzen}{1993}]{Mait93} Maitzen, H. M. 1993, \aaps, 102, 1
\bibitem[\protect\citeauthoryear{Maitzen \& Hensberge}{1981}]{Mait81} Maitzen, H. M., \& Hensberge, H. 1981, \aap, 96, 151
\bibitem[\protect\citeauthoryear{Maitzen \& Pavlovski}{1987}]{Mait87b} Maitzen, H. M., \& Pavlovski, K. 1987, \aaps, 71, 441
\bibitem[\protect\citeauthoryear{Maitzen \& Schneider}{1987}]{Mait87a} Maitzen, H. M., \& Schneider, H. 1987, \aaps, 71, 431
\bibitem[\protect\citeauthoryear{Maitzen \& Vogt}{1983}]{Mait83} Maitzen, H. M., \& Vogt, N. 1983, \aap, 123, 48
\bibitem[\protect\citeauthoryear{Maitzen et al.}{1997}]{Mait97} Maitzen, H. M., Paunzen, E., \& Rode, M. 1997, \aap, 327, 636
\bibitem[\protect\citeauthoryear{Maitzen et al.}{1998}]{Mait98} Maitzen, H. M., Pressberger, R., \& Paunzen, E. 1998, \aaps, 128, 573
\bibitem[\protect\citeauthoryear{Majaess et al.}{2007}]{Maja07} Majaess, D. J., Turner, D. G., \& Lane, D. J. 2007, \pasp, 119, 1349
\bibitem[\protect\citeauthoryear{Mathys et al.}{1989}]{Math89} Mathys, G., Maitzen, H. M., North, P., et al. 1989, The Messenger, 55, 41
\bibitem[\protect\citeauthoryear{Maury}{1897}]{Maur97} Maury, A. 1897, Ann. Astron. Obs. Harvard, vol. 28, Part 1
\bibitem[\protect\citeauthoryear{McCuskey}{1974}]{Mccu74} McCuskey, S. W. 1974, \aj, 79, 107
\bibitem[\protect\citeauthoryear{Mermilliod}{1982}]{Merm82} Mermilliod, J.-C. 1982, \aap, 109, 37
\bibitem[\protect\citeauthoryear{Mikul{\'a}{\v s}ek et al.}{2011}]{Miku11} Mikul{\'a}{\v s}ek, Z., Krti\v{c}ka, J., Henry, G. W., et al. 2011, \aap, 534, L5
\bibitem[\protect\citeauthoryear{Napiwotzki et al.}{1993}]{Napi93} Napiwotzki, R., Schoenberner, D., \& Wenske, V. 1993, \aap, 268, 653 
\bibitem[\protect\citeauthoryear{Netopil et al.}{2008}]{net08} Netopil, M., Paunzen, E., Maitzen, H. M., North, P., \& Hubrig, S. 2008, \aap, 491, 545 
\bibitem[\protect\citeauthoryear{Niedzielski \& Muciek}{1988}]{Nied88} Niedzielski, A., \& Muciek, M. 1988, \actaa, 38, 225
\bibitem[\protect\citeauthoryear{North}{1987}]{Nort87} North, P. 1987, \aaps, 69, 371
\bibitem[\protect\citeauthoryear{Paunzen \& Netopil}{2006}]{Paun06} Paunzen, E., \& Netopil, M. 2006, \mnras, 371, 1641
\bibitem[\protect\citeauthoryear{Paunzen et al.}{2005}]{Paun05} Paunzen, E., St{\"u}tz, Ch., \& Maitzen, H. M. 2005, \aap, 441, 631
\bibitem[\protect\citeauthoryear{Paunzen et al.}{2001}]{Paun01} Paunzen, E., Duffee, B., Heiter, U., Kuschnig, R., \& Weiss, W. W. 2001, \aap, 373, 625
\bibitem[\protect\citeauthoryear{Paunzen et al.}{2010}]{Paun10} Paunzen, E., Heiter, U., Netopil, M., \& Soubiran, C. 2010, \aap, 517, A32
\bibitem[\protect\citeauthoryear{Pavlovski \& Maitzen}{1989}]{Pavl89} Pavlovski, K., \& Maitzen, H. M. 1989, \aaps, 77, 351
\bibitem[\protect\citeauthoryear{P{\"o}hnl et al.}{2005}]{Poeh05} P{\"o}hnl, H., Paunzen, E., \& Maitzen, H. M. 2005, \aap, 441, 1111
\bibitem[\protect\citeauthoryear{Preston}{1974}]{Pres74} Preston, G. W. 1974, \araa, 12, 257
\bibitem[\protect\citeauthoryear{Renson \& Manfroid}{2009}]{Rens09} Renson, P., \& Manfroid, J. 2009, \aap, 498, 961
\bibitem[\protect\citeauthoryear{R{\"o}ser et al.}{2010}]{Roes10} R{\"o}ser, S., Demleitner, M., \& Schilbach, E. 2010, \aj, 139, 2440
\bibitem[\protect\citeauthoryear{Schaller et al.}{1992}]{Scha92} Schaller, G., Schaerer, D., Meynet, G., \& Maeder, A. 1992, \aaps, 96, 269
\bibitem[\protect\citeauthoryear{Schild \& Romanishin}{1976}]{Schi76} Schild, R., \& Romanishin, W. 1976, \apj, 204, 493
\bibitem[\protect\citeauthoryear{Schmidt-Kaler}{1982}]{Schm82} Schmidt-Kaler, Th. 1982, In: Landolt-B{\"o}rnstein New Series, group VI, vol. 2b, p. 453
\bibitem[\protect\citeauthoryear{Sharma et al.}{2006}]{Shar06} Sharma, S., Pandey, A. K., Ogur,a K., Mito, H., Tarusawa, K., \& Sagar, R. 2006, \aj, 132, 1669
\bibitem[\protect\citeauthoryear{Skiff}{2013}]{Skif13} Skiff, B. A. 2013, General Catalogue of Stellar Spectral Classifications, Lowell Observatory
\bibitem[\protect\citeauthoryear{Szklarski \& Arlt}{2013}]{Szkl13} Szklarski, J., \& Arlt, R. 2013, \aap, 550, A94
\bibitem[\protect\citeauthoryear{Tetzlaff et al.}{2011}]{Tetz11} Tetzlaff, N., Neuh{\"a}user, R., \& Hohle, M. M. 2011, \mnras, 410, 190
\bibitem[\protect\citeauthoryear{Voroshilov et al.}{1985}]{Voro85} Voroshilov, V. I., Guseva, N. G., Kalandadze, N. B., et al., 1985, Catalog of BV magnitudes 
and spectral classes of 6000 stars, Kiev, Izdatel'stvo Naukova Dumka  
\bibitem[\protect\citeauthoryear{Young \& Martin}{1973}]{Youn73} Young, A., \& Martin, A. E. 1973, \apj, 181, 805
\bibitem[\protect\citeauthoryear{Zacharias et al.}{2013}]{Zach13} Zacharias, N., Finch, C. T., Girard, T. M., et al. 2013, \aj, 145, 44
\bibitem[\protect\citeauthoryear{Zejda et al.}{2012}]{Zejd12} Zejda, M., Paunzen, E., Baumann, B., Mikul\'a\v sek, Z., \& Li\v ska, J. 2012, \aap, 548, A97
\bibitem[\protect\citeauthoryear{Zelwanowa}{1971}]{Zelw71} Zelwanowa, E. 1971, Astron. Nachr., 293, 163
\end{thebibliography}
\end{document}